\documentclass[structabstract]{aa}  
\usepackage{graphicx}
\usepackage{natbib}
\bibpunct{(}{)}{;}{a}{}{,} 
\usepackage{longtable,lscape}
\begin{document}
   \title{Angle-dependent radiative grain alignment}

   \subtitle{Confirmation of a magnetic field -- radiation anisotropy angle dependence on the efficiency of interstellar grain alignment}

   \author{B-G Andersson \inst{1}\thanks{Visiting Astronomer, Complejo Astron\'{o}mico El Leoncito, operated under agreement between the Consejo Nacional de Investigaciones Cient'ficas y T\'{e}cnicas de la Rep\'{u}blica Argentina and the National Universities of La Plata, C\'{o}rdoba and San Juan.}\fnmsep
          \and
          O. Pintado \inst{2}\fnmsep\inst{*}
          \and 
          S.B. Potter \inst{3}
          \and
          V. Strai\v{z}ys \inst{4}
          \and
          M. Charcos-Llorens \inst{1}
          }

   \institute{USRA/SOFIA Science Center, NASA Ames Research Center,
              Mail Stop N211-3, Moffett Field, CA 94035, USA\\
         \and 
Instituto Superior de Corellaci\'{o}n Geol\'{o}gica, CONICET, Miguel Lillo 205, 4000 San Miguel de Tucum\'{a}n, Argentina\\
         \and 
South African Astronomical Observatory, PO Box 9, Observatory 7935, Cape Town, South Africa\\
         \and 
Institute of Theoretical Physics and Astronomy, Vilnius University, Go\v{s}tauto 12, Vilnius LT-01108, Lithuania\\
             }

   \date{ }

 \abstract
{Interstellar grain alignment studies are currently experiencing a renaissance due to the development of a new quantitative theory based on Radiative Alignment Torques (RAT).  One of the distinguishing predictions of this theory is a dependence of the grain alignment efficiency on the relative angle  ($\Psi$) between the magnetic field and the anisotropy direction of the radiation field.  In an earlier study we found observational evidence for such an effect from observations of the polarization around the star HD~97300 in the Chamaeleon I cloud.  However, due to the large uncertainties in the measured visual extinctions, the result was uncertain.}
{By acquiring explicit spectral classification of the polarization targets, we have sought to perform a more precise reanalysis of the existing polarimetry data.}
{We have obtained new spectral types for the stars in our for our polarization sample, which we combine with photometric data from the literature to derive accurate visual extinctions for our sample of background field stars.  This allows a high accuracy test of the grain alignment efficiency as a function of $\Psi$.}
{We confirm and improve the measured accuracy of the variability of the grain alignment efficiency with $\Psi$, seen in the earlier study.  We note that the grain temperature (heating) also shows a dependence on $\Psi$ which we interpret as a natural effect of the projection of the grain surface to the illuminating radiation source.  This dependence also allows us to derive an estimate of the fraction of aligned grains in the cloud.}
{} 
   \keywords{Radiation mechanisms: general, Techniques: polarimetric, (ISM:) dust, extinction, ISM: magnetic fields}

   \maketitle
%

\section{Introduction}
  Broadband optical/infrared interstellar polarization was first detected in 1949 \citep{hall1949, hiltner1949a, hiltner1949b} and was already from the start assumed to be associated with dichroic extinction due to asymmetric dust grains aligned with their long axis across the magnetic field direction \citep{hiltner1949b, spitzer1949}.  However, despite over 60 years of efforts, the details of the grain alignment process are still not fully understood.   A quantitative understanding of the physical processes responsible for the grain alignment and hence polarization will allow us not only a better understanding of interstellar magnetic fields, by e.g. putting the Chandrasekhar-Fermi method \citep{chandrasekhar1953} on a more solid footing, but will likely also provide new probes of the characteristics of the dust.

The long standing ''text book'' explanation for the alignment, via paramagnetic relaxation in rotating grains, was put forward in the seminal paper by Davis and Greenstein (\citealt{davis1951}: DG).  Over the following decades, various modifications were then proposed to enhance the efficiency of this mechanism, including proposals for enhancing the magnetic susceptibility of the material \citep{jones1967, mathis1986} and enhanced torques on the grains due to particle ejections from the grain surface \citep{purcell1979}.  Generally, the driving mechanism for the grain spin-up in the various modifications to the DG mechanism are fixed in the grain's coordinate system (including the above so called "Purcell rockets" if the particle ejection sites are restricted to specific locations on the grain surface).  

Another long-standing possibility to produce grain alignment is through mechanical alignment in situations with relative motion between the gas and dust \citep{gold1952,lazarian1996a}.  This flow will tend to cause the grains to align themselves with as small a collisional cross-section as possible to the flow \citep{lazarian1996a}.  Because most large scale flows in the ISM (including winds from hot stars) tend to be ionized, these flows are usually constrained to be along the magnetic field lines and therefore cause grain alignment with the long axis of the grain along the magnetic field direction and polarization perpendicular to the magnetic field direction.

Starting in the 1990s several authors, including \citet{lazarian1999, lazarian1999b} and \citet{roberge2004} put the paramagnetic relaxation paradigm into doubt by showing that, due to the internal energy modes of the grains, a dust grain will tend to change its orientation in space on relatively short time scales.  Since such a "flip" will cause a torque fixed in the grain to drive the grain rotation in the opposite direction (in a space-based coordinate system) the grain never achieves a significant angular momentum, making paramagnetic relaxation alignment inefficient.  

Parallel to the discovery of these challenges for the classical DG model, an alternative theory of interstellar grain alignment was being worked out by Bruce Draine, Alexander Lazarian and their coworkers (e.g. \citealt{draine1997,lazarian2007}).  Based on early work by \citet{dolginov1976} they showed that an irregular grain with a net helicity will be spun up by the differential scattering of the left and right-hand circular components in an external light source.  Over many periods of Lamour precession, the light can then also align the grain with their long axis perpendicular to the magnetic field, without any contribution from paramagnetic relaxation.   Because helicity is invariant on reflection, this mechanism is not susceptible to the limitations by the "thermal flipping" of the grain.  The sole environmental requirement for alignment by this mechanism is that the radiation field be anisotropic.  In the interstellar medium, this is almost always the case.  Hence, while the physical mechanisms responsible for the grain alignment are quite distinct, to first order extended DG alignment and radiative alignment provide the same observational prediction; namely polarization parallel to the projected magnetic field direction. 

Radiative Alignment Torque (RAT) theory has matured over the last decade and is now providing a number of specific, testable, predictions \citep{lazarian2007}.  One such prediction, which is especially attractive for probing the validity of the theory, since it is unique to RAT theory, is that the alignment efficiency, but not the polarization angle, should vary with the angle between the magnetic field and the radiation field anisotropy ($\Psi$; see \citealt{lazarian2007}).   One way to observationally probe this prediction is to find dust grains for which the dominant radiation source is a localized source (a star) and measure the alignment efficiency for background stars projected at different position angles around the source star.  Figure 1 illustrates the geometry of the situation.

   \begin{figure}
   \centering
   \includegraphics[width=8cm]{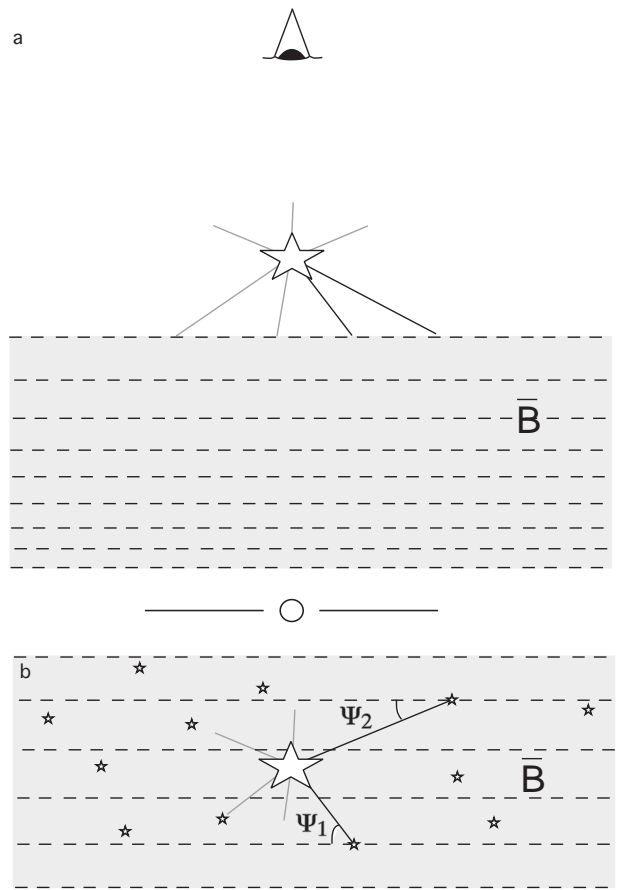}
      \caption{The geometry used to probe the effects of the relative orientation of the magnetic field and radiation field anisotropy is illustrated.  In the two panels, the dashed lines illustrate an idealized magnetic field. (a; side view) A bright source is located close to and in front of an underlying interstellar cloud surface, such that its radiation locally dominates over the diffuse radiation field.  (b; \textit{en face} view)  By probing the polarization of background field stars we can probe the effects of angle dependent Radiative Alignment Torque (RAT) grain alignment (Adapted, with permission of the AAS, from \citealt{bga2010})}
         \label{cartoon}
   \end{figure}
We found such a combination of star and nearby dust in the Chamaeleon I cloud and showed in \citet{bga2010}, combining new polarimetry with the multi-band polarimetry from \citet{whittet1992}, that the grain alignment is indeed enhanced at small projected distances to the central source star (in this case HD~97300).  A preliminary study of the angle dependence of the alignment was also carried out in \citet{bga2010}, but was limited by the fact that, for most of our field stars, the spectral classification and therefore the determinations of the visual extinctions on each line of sight, had to be accomplished using near infrared (NIR; 2MASS) photometry alone.   Because the small color excesses in the NIR and uncertainties in the reddening vector slope, the spectral classifications typically had uncertainties of a full spectral class, translating to uncertainties on the visual extinctions of $\sim$1 \textrm{mag}.
Despite these limitations, we detected a relative grain alignment enhancement for position angles where the magnetic field and radiation field anisotropy were parallel, in agreement with the predictions of RAT theory.

In order to test whether the large uncertainties in the derived visual extinctions confused the analysis, we here present a reanalysis of the polarization data from \citet{bga2010}, using new explicit spectral classifications of the field star acquired at the the "Complejo Astronomico de Leoncito" (CASLEO) in Argentina.


\section{Observations, data reduction and analysis}
We used the REOSC spectrometer \citep{pintado1996} on the 2.15~m telescope at the CASLEO on the nights of 2011, March 4--6.  The spectrometer was used with the 300 lines/mm grating, producing a measure wavelength coverage of $\lambda\lambda$3631--7128~$\textrm{\AA}$ with a spectral resolution of $\Delta\lambda$=3.4~$\textrm{\AA}$/pixel.  The detector used was a thinned TEK 1024$\times$1024 CCD with 24~$\mu$m pixels.  Wavelength calibration was achieved using exposures of Thorium-Argon hollow cathode lamps.  No attempt was made to achieve photometric calibration of the data as the sky conditions did not warrant it.

The data reduction was achieved using standard procedures and routines in the IRAF environment.  After bias and flat-field corrections, the two-dimensional spectra were traced and integrated across the dispersion direction and extracted into one-dimensional form.  The extracted spectra were then wavelength calibrated, normalized and compared, interactively, to the standard sequence from \citet{jacoby1984}.  The classifications were redone independently several times, by two of the authors.   The resulting spectral classifications are listed in Table 1.  Also listed in Table 1 are the derived colour excesses, based on Tycho \citep{tycho2000} photometry and intrinsic colours from \citet{cox2000}.  In a small number of cases Tycho photometry is not available, and in those cases we have instead used NOMAD \citep{zacharias2005} photometry.  Visual extinctions ($A_V$) were calculated using total-to-selective extinction ($R_V$) values derived, from $A_V=R_V \times E_{B-V}$, using the relation: $R_V=1.1 \times E_{V-K}/E_{B-V}$ from \citet{whittet1978}.

As discussed in \citet{bga2007}, the wavelength of the maximum of the polarization curve ($\lambda_{\rm max}$) is a sensitive probe of the grain alignment and one that is immune to the magnetic field topology along the line of sight.  The absolute value of $\lambda_{\rm max}$ depends on the over-all grain size distributions and likely on the color and intensity of the local diffuse interstellar radiation field (ISRF), but as shown by \citet{bga2007} the slope in the $\lambda_{\rm max}$ vs. $A_V$ relation is universal.  Hence we can measure the relative grain alignment enhancement in a region by deriving the average relationship between $\lambda_{\rm max}$ and $A_V$ for field stars behind the cloud in question and then probing for localized deviations from this average relationship.  We did this in \citet{bga2007,bga2010} and found that in Chamaeleon I, but beyond the region around HD~97300 (where its radiation field dominates that of the diffuse radiation), a linear relationship is found:

\begin{equation}
\langle\lambda_{\rm max}(A_V)\rangle=(0.527\pm0.006)+(0.033\pm0.003)\times A_V.
\end{equation}

\section{Results and discussion}
  
\subsection{Stellar classification}
The uncertainties for the spectral classifications are based on estimates from the two independent classifiers.  Usually the results by the individual classifiers agreed to within the assigned uncertainties.  The luminosity class assignments are less certain.  For early-type star (up to F) we assume that the stars are on the Main Sequence.  For later stellar types (G and beyond), we used the known distance to the Chamaeleon I cloud ($d=150\pm 15$~pc, \citealt{whittet1997}), spectroscopic parallaxes for luminosity classes III and V, and the measured colour excesses, derived for each luminosity class, to set the luminosity class.  The restriction to these two luminosity classes could be a source of some error, in a few cases where luminosity classes IV (or I) might have been more appropriate.   Specifically, TYC 9410-1931-1 and TYC 9414-0046-1 have been assigned a spectral class of G4 III.  However, this spectral/luminosity class is located in the Herzsprung gap of the HR diagram where the number of stars is quite small.  As noted below some stars show colour excess ratios that would indicate that they are part of binaries or show the effects of peculiar circumstellar material, or other effects yielding non-standard colours.  The former of these two stars falls in this category.

For five stars in our sample the colour excess $E_{B-V}$ is consistent with zero and we have excluded these stars from the subsequent analysis.
Similarly, for six stars the colour excess ratio $E_{V-K}$/$E_{B-V}$ is significantly smaller than the nominal value 2.74, assuming a standard interstellar extinction curve and a $R_V$ value of 3.1 \citep{cox2000}.  After checking the spectral classifications for these stars we conclude that the discrepancy is in the measured colours and may indicate binarity, or other non-standard spectral behaviour.  We have therefore excluded also these stars from the analysis of the grain alignment variations.

While a systematic increase in $R_V$ is expected for grain growth, no clear correlation is seen between $R_V$ and column density (e.g. $I_{100}$).  This is likely because the total column densities in our sample are still fairly small.

\subsection{Grain alignment}
\begin{figure}
   \centering
   \includegraphics[width=8cm]{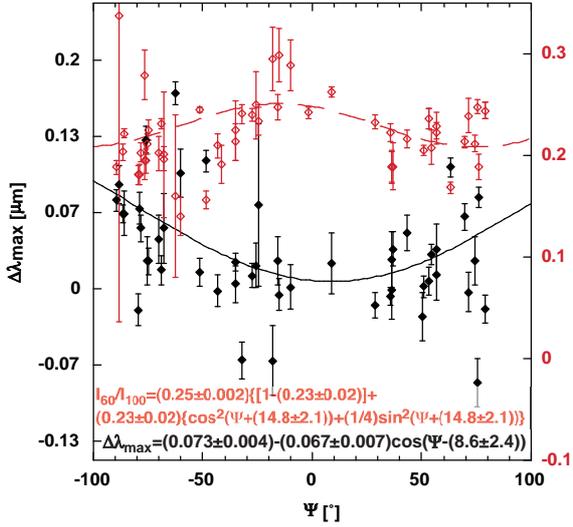}
      \caption{The variations in grain alignment efficiency, as probed by the location of the peak in the polarization curve, is shown as a function of the relative angle between the magnetic field and radiation field anisotropy: $\Psi$ (filled, black, diamonds.  Also shown are the variations in the FIR ratio $I_{60}/I_{100}$ (open, red, diamonds).  Both show a variation with $\Psi$.  The change in the grain temperature can be understood in terms of the surface area facing the illuminating star, while the variations in $\lambda_{\rm max}$ are consistent with predictions by RAT theory.}
         \label{results}
   \end{figure}

Figure \ref{results} shows the relative grain alignment efficiencies for the stars in our sample as measured by the above technique in filled (black) diamonds as a function of $\Psi$ (see Figure \ref{cartoon}):

\begin{equation}
\Delta\lambda_{\rm max}^i=\lambda_{\rm max}^i-\langle\lambda_{\rm max}(A_V^i)\rangle.
\end{equation}

The FIR ratio I$_{60}$/I$_{100}$ is also shown, in open (red) diamonds.  The FIR data here are from the IRIS reprocessing of the IRAS data \citep{IRIS2005}.  For details see \citet{bga2007}.

In Figure \ref{results} we have over-plotted the best fit of the function $\Delta\lambda_{\rm max}=A+B \times \textrm{cos}(\Psi-\Psi_0)$ where $\Psi$ is the position angle from the star, relative to the average magnetic field direction in the area, as determined from the polarization maps by \citet{mcgregor1994, bga2010} and $\Psi_0$ is the symmetry angle of the function (in this case the minimum).  We have chosen a simple cosine relation here, since the exact theoretical functional form for the grain alignment depends on several unknown parameters of the grains, the radiation field and the structure of the cloud along the line of sight.  With the new and significantly improved visual extinctions for our background stars, we find a statistically significant depression in $\Delta\lambda_{\rm max}$ $\Psi_0$ close to 0, in agreement with theoretical predictions.  An F-test \citep{lupton1993} yields a greater than 99\% probability that the two additional parameters in the cosine function are statistically warranted (as compare to a simple average value).  The parameters in the current best fit are all within the mutual uncertainties of the best fit in \citet{bga2010}.   We now find a 9.5$\sigma$ deviation from a null result for the amplitude of the grain alignment enhancement.

\begin{figure}
  \centering
\includegraphics[width=8cm]{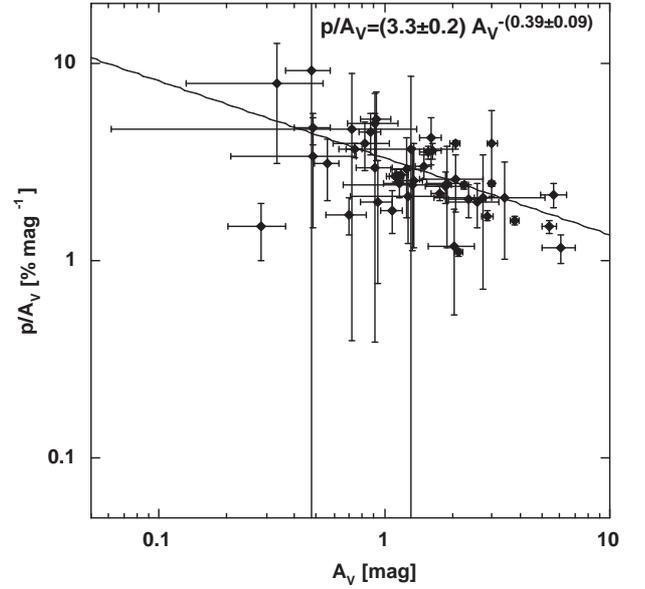}
\caption{The fractional polarization (p$_{\rm max}$/$A_V$) for the sample is plotted as a function of the visual extinction.  The power-law exponent is consistent, within the mutual error-bars, with the one found for deeper lines of sight by \citet{whittet2008}.}
\label{pdAv}%
   \end{figure}

The fractional polarization ($p_{\rm max}/A_V$; Fig \ref{pdAv}) for our combined sample shows a power-law slope $b=-0.39\pm0.09$ consistent, within 1.2$\sigma$ mutual uncertainty, from the results in \citet{whittet2008}: $b=-0.52\pm0.07$.

As discussed in \citet{bga2010}, while we do detect an alignment efficiency enhancement in the direction of the magnetic field, relative to the star, we do not see any evidence for a rotation of the polarization angles in this direction, as would be expected for mechanical alignment (see Figure 9 of \citealt{bga2010}).

\subsection{Grain heating}
\begin{figure}
  \centering
\includegraphics[width=8cm]{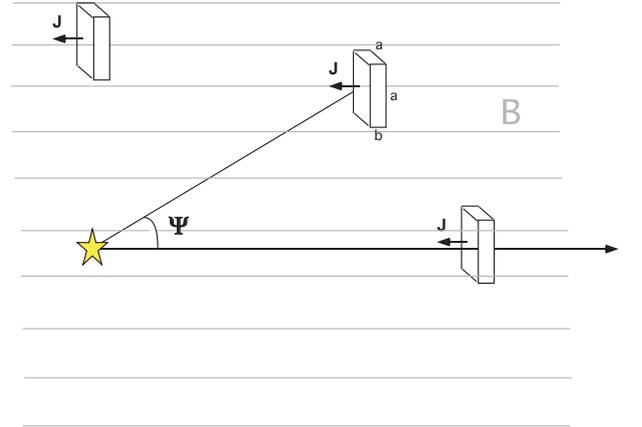}
\caption{The heating of asymmetrical grains, aligned with the magnetic field (thin, gray, lines) will vary with the angle $\Psi$.  If only a fraction of the grains are alignable, we can use the variations in the dust temperature to estimate the fraction of aligned grains or the axes ratio of these asymmetric grains}
\label{heatingcartoon}%
   \end{figure}

Figure \ref{results} also shows a systematic variation in the $I_{60}/I_{100}$ ratio as a function of $\Psi$.  To model this we envision the alignable grains as a system of simple parallelepipeds with two large sides of length $a$, and a smaller side of length $b$ (like a pizza box; Figure \ref{heatingcartoon}).  The radiative heating of these grains around the star then takes the form:
\begin{equation}
I\propto\frac{I_0}{r^2}a^2[\textrm{cos}^2(\Psi)+\frac{b}{a}\textrm{sin}^2(\Psi)],
\end{equation}

\noindent where $I_0$ is the effective intensity of the central star and $r$ is  the distance between the star an dust grain.  Figure \ref{heatingcartoon} illustrates the model geometry.  We also need to allow for symmetrical grains that will not show a dependence on the angle $\Psi$.  As shown by \citet{desert1990} the $I_{60}/I_{100}$ ratio shows a roughly linear response to the radiation field strength in the range 1-5 times the local interstellar radiation field.  Hence, as a simple first order, heuristical, model we can fit the FIR colour temperature to the function:

\begin{equation}
\frac{I_{60}}{I_{100}}(\Psi)=c_1[(1-\epsilon)+\epsilon(\textrm{cos}^2(\Psi)+(b/a)\textrm{sin}^2(\Psi))],
\label{fitequ}
\end{equation}

\noindent where $\epsilon$ is the fraction of aligned, asymmetrical grains and $b/a$ is the axes ratio of the asymmetrical grains.   

The two parameters $\epsilon$ and ($b/a$) are degenetrate in the fits.  We therefore ran a number of fits using $\epsilon$ as a free parameter with $(b/a)^{-1}$ set to $\surd$2, 2, 4 and 6.  For these choices of ($b/a$) we find $\epsilon$=(0.58$\pm$0.04), (0.34$\pm$0.02), (0.23$\pm$0.02) and (0.20$\pm$0.02).  Figure \ref{results} shows the best fit for $(b/a)^{-1}$=4.

\citet{kim1995} present, in their Table 1, the polarization-to-extinction ratio for perfectly aligned oblate spheroids with the axes ratios we used in our fits.  They also quote the maximum of the measured value of this ratio as $p_{\rm max}/\tau$=0.028.  If we use the ratio of the observed to calculated value in their paper as signifying the required dilution of the aligned oblate spheroidals (by unaligned, or fully symmetrical grains), we find that the fractions of aligned grains they predict are 0.56, 0.31, 0.18 and 0.17, very close to the best fit values in our model.

The fitting constant $c_1$ in equation \ref{fitequ} contains the distance from the illuminating star to each line of sight.  Our sample is not large enough to conclusively show this dependence for different distance annuli, but splitting the sample in two and fitting the targets nearest to HD~97300 does produce an increased value of $c_1$, compared both to the full sample and the more distant half-sample.

Repeating this analysis on the full maps of IRAS data in the Chamaeleon region produces consistent results albeit with larger scatter in the FIR ratio as a function of position angle (in particular, for the annuli $d$=0.3--0.4~deg and $d$=0.4--0.5~deg, the best fit values for $\epsilon$ agree, within the error bars, with the ones found for our polarimetry star sample).  Further analysis, in particular repeating the experiment for other cloud regions dominated by a near-by star, will be needed to ascertain whether this effect is real or whether it is a statistical aberration in the current data set.

\section{Conclusions}

With the improved observational data, we confirm the preliminary result from \citet{bga2010} that the grain alignment seems to depend on $\Psi$, as predicted by RAT theory.  This provides an important observational constraint on, and in this case support for, the currently best developed theory for interstellar grain alignment.  With further supportive observational tests of the theory, the second-longest standing mystery\footnote{The longest-standing mystery of ISM astrophysics is the nature of the carriers of the Diffuse Interstellar Bands, detected \citep[see][]{friedman2011} in 1919 \citep{heger1922} and identified as being of interstellar origin in 1936 \citep{merrill1936}} of interstellar medium astrophysics may be within reach of resolution.

We also find a dependence on $\Psi$ in the $I_{60}/I_{100}$ ratio. Since the aligned grains causing the polarization have their major axis perpendicular to the magnetic field, this enhancement in the FIR ratio should be expected simply as a projection effect of the grain surface to the radiation from the central star.  Our best fit models of the FIR ratio find that the fraction of aligned grains is very close to that found for theoretical models of the extinction and polarization curves by \citet{kim1995}.  

It is possible that an underlying corrugation of the cloud surface could partly be responsible for the $\Psi$ dependence.  This could happen if a ridge centered on HD~97300 is oriented parallel to the average magnetic field.  However, the low estimated magnetic field of the Chamaeleon I cloud, $B_\parallel=3\pm4~\mu G$ \citep{bourke2001} makes this unlikely.  Repeating the experiment in other regions where a localized radiation source dominates the grain illumination will allow such caveats to be addressed.

%
%
%
\begin{landscape}
\begin{table*}
\caption{\label{t7}Spectral types and photometry for stars in the region.}
\centering
\begin{tabular}{lrrrrlcll}
\hline\hline
ID              & RA          & Dec          & $V$              & $B$--$V$           & Sp. Class\tablefootmark{a} & Source\tablefootmark{b} & $R_V$\tablefootmark{c}  & $A_V$\\
                & (2000)      & (2000)       & [mag]         & [mag]         &          &              &        & [mag]\\
\hline

TYC 9414-0186-1 & 10:59:06.97 & -77:01:40.30 & 11.66$\pm$0.15 & 1.26$\pm$0.32 & G9 V (2) & A & 5.58$\pm$0.72 & 2.76$\pm$0.26\\
TYC 9410-1931-1 & 10:59:29.16 & -76:11:15.50 & 11.29$\pm$0.11 & 1.72$\pm$0.32 & G4 III (1) & A & 0.86$\pm$0.45 & 0.77$\pm$0.14\\
TYC 9410-2587-1 & 10:59:39.81 & -76:24:15.60 & 11.17$\pm$0.08 & 0.51$\pm$0.10 & G2 V (1) & A & --- & --- \\
TYC 9410-0532-1 & 11:01:25.94 & -76:45:07.40 & 9.83$\pm$0.03 & 0.47$\pm$0.04 & A9 V (1) & A & 4.43$\pm$0.29 & 0.87$\pm$0.09\\
HD 95883        & 11:01:28.86 & -75:39:51.98 & 7.33$\pm$0.01 & 0.13$\pm$0.02 & A2 Vn... (1) & B & 3.55$\pm$0.47 & 0.28$\pm$0.08\\
TYC 9414-0642-1 & 11:03:11.57 & -77:21:04.28 & 11.48$\pm$0.11 & 1.11$\pm$0.22 & G2 V (1) & A & 6.32$\pm$2.96 & 3.01$\pm$0.17\\
TYC 9414-0150-1 & 11:03:31.68 & -76:56:04.89 & 10.69$\pm$0.05 & 0.37$\pm$0.06 & A2 (2) & C & 4.97$\pm$0.32 & 1.61$\pm$0.17\\
TYC 9410-2633-1 & 11:03:36.31 & -75:59:03.40 & 10.76$\pm$0.07 & 0.55$\pm$0.09 & G4 V (2) & A & 0.18$\pm$6.15 & $<$0.09\\
TYC 9410-2362-1 & 11:04:10.22 & -76:22:18.20 & 10.89$\pm$0.06 & 0.73$\pm$0.09 & A7 V (2) & A & 1.30$\pm$0.36 & 0.68$\pm$0.17\\
TYC 9410-2504-1 & 11:04:37.92 & -76:19:00.90 & 11.37$\pm$0.10 & 0.63$\pm$0.15 & F8 V (1) & A & 9.22$\pm$1.53 & 1.03$\pm$0.13\\
TYC 9410-2172-1 & 11:04:43.80 & -76:16:45.20 & 11.47$\pm$0.11 & 0.90$\pm$0.21 & F5 V (2) & A & 1.21$\pm$0.76 & 0.55$\pm$0.28\\
TYC 9410-1907-1 & 11:06:07.52 & -75:55:09.10 & 10.58$\pm$0.05 & 0.94$\pm$0.08 & F8 III (1) & A & 2.72$\pm$0.30 & 1.09$\pm$0.12\\
TYC 9410-2117-1 & 11:08:23.42 & -75:56:34.20 & 11.12$\pm$0.08 & 1.03$\pm$0.16 & M2 V (1) & A & --- & --- \\
N0136-0067899   & 11:09:07.70 & -76:18:14.80 & 11.87$\pm$0.10 & 1.10$\pm$0.10 & K7 III (2) & A & --- & --- \\
TYC 9410-2426-1 & 11:12:50.73 & -75:58:07.00 & 9.82$\pm$0.03 & 0.48$\pm$0.03 & F0 (2) & C & 5.02$\pm$0.45 & 0.93$\pm$0.14\\
PPM 371023      & 11:12:54.83 & -75:52:11.93 & 9.15$\pm$0.02 & 1.40$\pm$0.04 & K5 (2) & C & 2.25$\pm$4.14 & $<$0.44\\
TYC 9410-2449-1 & 11:13:26.29 & -76:19:23.30 & 11.60$\pm$0.12 & 0.61$\pm$0.17 & G2 V (2) & A & --- & --- \\
TYC 9414-0649-1 & 11:13:48.08 & -77:50:12.37 & 11.59$\pm$0.14 & 1.26$\pm$0.30 & G7 III (1) & A & 2.62$\pm$0.98 & 0.91$\pm$0.17\\
TYC 9414-0462-1 & 11:14:09.38 & -77:14:49.50 & 10.93$\pm$0.08 & 1.85$\pm$0.29 & K3 III (2) & A & 3.30$\pm$0.62 & 1.90$\pm$0.36\\
TYC 9414-0260-1 & 11:14:11.23 & -77:01:15.10 & 10.96$\pm$0.07 & 0.51$\pm$0.08 & F2V (2) & A & 8.68$\pm$0.77 & 1.32$\pm$0.22\\
TYC 9414-0392-1 & 11:14:29.64 & -77:07:06.40 & 11.19$\pm$0.09 & 0.73$\pm$0.13 & G2 V (1) & A & 4.64$\pm$1.46 & 0.47$\pm$0.11\\
TYC 9414-0046-1 & 11:14:45.15 & -76:58:20.80 & 11.33$\pm$0.09 & 1.22$\pm$0.21 & G4 III (1) & A & 3.44$\pm$0.61 & 1.35$\pm$0.14\\
TYC 9410-1825-1 & 11:15:17.32 & -76:08:56.70 & 11.70$\pm$0.12 & 0.30$\pm$0.14 & B9 V (1) & A & 3.62$\pm$0.51 & 1.26$\pm$0.19\\
TYC 9410-2634-1 & 11:15:30.57 & -76:29:02.30 & 11.68$\pm$0.14 & 1.37$\pm$0.30 & F8 V (1) & A & 1.01$\pm$0.44 & 0.86$\pm$0.17\\
TYC 9410-2731-1 & 11:15:57.31 & -76:29:08.50 & 11.27$\pm$0.10 & 2.23$\pm$0.37 & M0 III (2) & A & 3.04$\pm$0.65 & 2.04$\pm$0.47\\
TYC 9410-2612-1 & 11:17:30.79 & -76:27:07.90 & 10.07$\pm$0.04 & 2.03$\pm$0.17 & M3 III (1) & A & 3.02$\pm$0.67 & 1.27$\pm$0.56\\
TYC 9414-0080-1 & 11:18:19.91 & -77:07:16.70 & 9.98$\pm$0.03 & 1.10$\pm$0.06 & G7 III (2) & A & 2.99$\pm$0.40 & 0.56$\pm$0.07\\
TYC 9410-0488-1 & 11:18:21.19 & -76:40:56.00 & 10.80$\pm$0.07 & 2.27$\pm$0.29 & K6 III (1) & A & 1.22$\pm$0.46 & 0.91$\pm$0.13\\
TYC 9410-1231-1 & 11:19:54.76 & -75:38:38.21 & 10.27$\pm$0.04 & 0.52$\pm$0.05 & A5 (2) & C & 1.87$\pm$0.32 & 0.69$\pm$0.14\\
TYC 9415-0403-1 & 11:20:07.20 & -76:55:24.00 & 11.69$\pm$0.12 & 0.43$\pm$0.15 & A6 V (1) & A & 3.72$\pm$0.70 & 0.94$\pm$0.15\\
TYC 9411-0934-1 & 11:20:47.70 & -76:37:11.30 & 11.73$\pm$0.15 & 1.69$\pm$0.35 & K4 III (2) & A & 4.19$\pm$1.57 & 1.32$\pm$0.69\\
TYC 9411-1630-1 & 11:21:50.85 & -76:37:25.70 & 10.38$\pm$0.05 & 1.92$\pm$0.19 & K4 III (2) & A & 2.44$\pm$0.81 & 1.32$\pm$0.67\\
HD 99418    & 11:24:51.67 & -76:36:42.59 & 9.22$\pm$0.02 & 1.32$\pm$0.04 & G9 III: (1) & B & 4.26$\pm$0.18 & 1.49$\pm$0.11\\
HD 99558        & 11:25:47.95 & -76:30:29.15 & 9.13$\pm$0.02 & 1.19$\pm$0.03 & K0 III (1) & B & 4.90$\pm$0.53 & 0.91$\pm$0.23\\
HD 99591        & 11:26:04.34 & -75:43:34.95 & 9.37$\pm$0.02 & 0.54$\pm$0.03 & F7 V (2) & C & 0.51$\pm$10.18 & 0.03$\pm$0.25\\
\hline
\end{tabular}
\tablefoottext{a}{The number in parenthesis gives the estimated uncertainty of the spectral class, in sub-classes};
\tablefoottext{b}{Source of spectral classification: A -- This work, B -- Michigan Spectral Survey \citep{mss1999}, C -- SIMBAD database};
\tablefoottext{c}{In a few cases unphysical results were obtained for $R_V$.  These likely indicate non-standard spectral classes.  Those values have been omitted and marked by --- and the resulting visual extinctions were not used in the analysis.}
\end{table*}
\end{landscape}

\begin{acknowledgements}
We gratefully acknowledge the time allocation and support of the staff of the CASLEO observatory.  OIP contribution to this paper was partially supported by PIP0348 by CONICET.

We are grateful to Bill Reach for helpful discussions about the FIR response to grain heating as well as the anonymous referee and the A\&A editor for helpful suggestions improving the clarity of the manuscript.

\end{acknowledgements}

\bibliography{bgbiblio_tot}
\bibliographystyle{aa}


\end{document}